\documentclass[12pt,a4paper]{article}%
\usepackage{amsmath,amssymb,amsfonts}
\usepackage{graphicx,epsfig}
\usepackage{hyperref}
\usepackage{color}
\usepackage{amsmath}
\usepackage{amsfonts}
\usepackage{amssymb}
\usepackage{graphicx}%
\begin{document}
\begin{titlepage}
\begin{flushright}
UCB-PTH-06/04\\
LBNL-59894
\end{flushright}
\vskip 1.0cm
\begin{center}
{\Large \bf Improved Naturalness with a Heavy Higgs: \\An Alternative Road to LHC Physics}
\vskip 1.0cm
{\large Riccardo Barbieri$^a$, Lawrence J.~Hall$^b$ and Vyacheslav S.~Rychkov$^a$}\\[1cm]
{\it $^a$ Scuola Normale Superiore and INFN, Piazza dei Cavalieri 7, I-56126 Pisa, Italy} \\[5mm]
{\it $^b$ Department of Physics, University of California, Berkeley, and\\
Theoretical Physics Group, LBNL, Berkeley, CA 94720, USA}\\[5mm]
\vskip 1.0cm
\abstract{The quadratic divergences of the Higgs mass may be cancelled either
accidentally or by the exchange of some new particles.
Alternatively its impact on naturalness may be weakened by raising
the Higgs mass, which requires changing the Standard Model below its natural
cut-off. We show in detail how this can be achieved, while
preserving perturbativity and consistency with the electroweak
precision tests, by extending the Standard Model to include a
second Higgs doublet that has neither a vev nor couplings to
quarks and leptons. This Inert Doublet Model yields a perturbative
and completely natural description of electroweak physics at all
energies up to 1.5 TeV. The discrete symmetry that yields the
Inert Doublet is unbroken, so that Dark Matter may be composed of
neutral inert Higgs bosons, which may have escaped detection at
LEP2. Predictions are given for multilepton events with missing
transverse energy at the Large Hadron Collider, and for the direct detection of dark
matter. }
\end{center}
\end{titlepage}



\section{Introduction}

Unification, likely supersymmetric, as developed in the seventies and
eighties, is the most appealing and coherent picture that we have for physics
beyond the Standard Model. Clear experimental evidence for it would represent
a major breakthrough in physics and could orient the search for further
informative signals. Yet the current situation is ambiguous. From the
experimental viewpoint, on the positive side one has the unification of gauge
couplings and, with less numerical significance, the size of the neutrino
masses. On the negative side, however, one must consider the failure to find
any supersymmetric particle, any non-SM effect in flavour physics, any
evidence of proton decay, or, finally, a light Higgs boson. While there are
many explanations for the absence of these signals so far, and searches for
these phenomena should and will continue, we find it justified to consider
possible alternative roads for physics beyond the SM. Especially, although not
only, in these respects, the Large Hadron Collider should play a crucial role.
This is the general background behind this work.

Our expectations for the LHC are based on two observations: 1) the Higgs mass
gets quadratically divergent contributions, the dominant one being due to
virtual top quarks, which become comparable to the physical mass for the
cutoff $\Lambda_{t}\lesssim3.5 \; m_{h},$ and need to be cancelled by new
physics to avoid unnatural fine-tuning. 2) ElectroWeak Precision Tests (EWPT)
indicate that the SM Higgs is light, $m_{h}<186$ GeV at 95\% CL \cite{LEPEWWG}%
, with a central value considerably below the lower bound of 114 GeV from
direct searches. From 1) and 2) the standard view emerges, that the
divergence-cancelling physics, whatever it is (supersymmetry, Little Higgs,
\ldots) should be accessible at the LHC.

In this paper we consider the alternative possibility that the Higgs is heavy,
say $500$ GeV. In this case, the above conclusion does not apply, since the
naturalness cutoff from 1) is now raised to $\sim1.5$ TeV. Instead of
focussing on the new physics which cancels the top quark divergence (squarks,
vector quarks, \ldots), we must consider the modified electroweak theory below
$\Lambda_{t}$, that allows the heavy Higgs to pass the EWPT. Admittedly, to
guess which physics may render a Heavy Higgs compatible with the EWPT is not
easy. Some examples exist in the literature, starting from the work of
Einhorn, Jones and Veltman\cite{Einhorn:1981cy} and recently reviewed by
Peskin and Wells \cite{Peskin}. We postpone a few comments on this until
Section 4. Here we argue that the most obvious way to do this, while keeping
both naturalness and perturbativity, may reside in introducing an Inert
Doublet (ID) scalar, i.e.~a second Higgs without a vev or couplings to matter.

In the ID Model (IDM), the spectrum of the scalars, other than the true Higgs,
of mass $m_{h}$, consists of a charged state, of mass $m_{H}$, and of two
neutral states, of mass $m_{\text{L}}$ (L for lightest) and $m_{\text{NL}}$
(NL for next-to-lightest). The relation between these masses imposed by the
EWPT is fully analogous to the one that relates the Higgs mass and the Z mass
in the SM. In the entire perturbative regime of the IDM, we find that the
range of the radiative correction effects has a large overlap with the
corrections required to fit the precision data. We claim therefore that these
data do not prefer the light Higgs of the SM over the Heavy Higgs of the IDM.
On the other hand, in the IDM it is possible to raise the naturalness cut-off
to about $1.5$ TeV without fine tunings. Other than consequences for the LHC
mentioned above, this certainly ameliorates the problem posed by the
\textquotedblleft LEP paradox"\cite{paradox}, reducing by one order of
magnitude the fine tuning apparently needed to fix it. Indeed, this
improvement in naturalness is a major motivation for raising the Higgs boson
mass, and occurs more readily than in 2 Higgs doublet models with a light
Higgs \cite{BH}, and more simply than in ``Little Higgs'' models\footnote{A
possible connection between the fine tuning and the Higgs mass has also been
considered in ``Little Higgs'' models. See, e.g., Ref \cite{Gregoire:2003kr}%
.}. Furthermore, preliminary results from the TeVatron indicate a somewhat
lighter top quark, strengthening the upper bound on the SM Higgs mass, and
weakening the improved naturalness of the 2 Higgs doublet model with a light Higgs.

In most 2 Higgs doublet models a parity symmetry is introduced to ensure that
Higgs exchange does not give too large flavour changing amplitudes. In the IDM
the parity acts only on the inert doublet, and ensures that the doublet is
inert. Unlike conventional 2 Higgs doublet models, the parity is not
spontaneously broken by doublet vevs, and hence the Lightest Inert Particle,
or LIP, is stable. In much of the parameter space the LIP contributes only a
small fraction to the Dark Matter of the universe. But if there is a mild
degree of cancellation in the LIP mass, so that it is in the range of 70 GeV,
all DM can be accounted for by a neutral inert Higgs boson.

In Section 2 we set the stage by revisiting the SM with a heavy Higgs, paying
special attention to the improved naturalness, the triviality bound on the
Higgs mass, and to its incompatibility with EWPT constraints. In Section 3 we
present the IDM, discussing in detail all the analogous constraints. In the
same section we give a first description of the LHC signals of the IDM and we
discuss the properties of the LIP as a Dark Matter candidate. In Section 4 we
make a few comments on alternative models to render a heavy Higgs compatible
with the EWPT. Summary and Conclusions are given in Section 5.

\section{Standard Model with a Heavy Higgs}

\label{SM}

\subsection{Improved naturalness}

\label{SM.natur}

The SM is unnatural as a fundamental theory: the quadratic divergence of the
Higgs mass makes the electroweak scale highly sensitive to the UV cutoff.
Presumably, this quadratic divergence should be cancelled in the theory that
extends the SM to higher energy scales. Two known mechanisms for accomplishing
this are supersymmetry and realizing the Higgs as a pseudo-Goldstone boson.
Searching for such mechanisms amounts to what we might call the
\textit{qualitative} use of the naturalness principle. However, the principle
has also its other, \textit{quantitative} side. Namely, it can be used to
predict the energy scale by which the divergence-cancelling physics is
expected to appear. Such a prediction follows from comparing the size of the
one-loop quadratic divergence to the physical mass.

The quadratic divergence is given by ($v=174$ GeV)%
\begin{equation}
\delta m_{h}^{2}=\alpha_{t}\Lambda_{t}^{2}+\alpha_{g}\Lambda_{g}^{2}%
+\alpha_{h}\Lambda_{h}^{2} \label{quad}%
\end{equation}
where
\begin{equation}
\alpha_{t}=\frac{3m_{t}^{2}}{4\pi^{2}v^{2}},\;\;\alpha_{g}=-\frac{6m_{W}%
^{2}+3m_{Z}^{2}}{16\pi^{2}v^{2}},\;\;\alpha_{h}=-\frac{3m_{h}^{2}}{16\pi
^{2}v^{2}} \label{alphat}%
\end{equation}
and $\Lambda_{i}$ are the cutoffs on the momenta of the virtual top quarks,
gauge bosons, and the Higgs itself. We keep these cutoffs separate, because
generally there is no reason to expect that the physics cancelling all three
divergences will appear at exactly the same scale. In a more fundamental
theory, the various $\Lambda_{i}$ may be correlated, but if we do not specify
the theory which extends the SM\ and cancels the quadratic divergences, the
relative weight of the various terms in (\ref{quad}) cannot be
determined\footnote{Lumping all terms in (\ref{quad}) together with a common
value of $\Lambda_{i}=\Lambda$, one arrives at the conclusion that the SM has
no 1-loop fine-tuning problem provided that the quadratic divergences in
(\ref{quad}) cancel, which occurs for $m_{h}\approx300$ GeV (the so-called
Veltman condition \cite{Veltman}). For the reasons mentioned, we do not accept
this argument.}.

Knowing (\ref{quad}), we can compute the sensitivity of the Higgs mass to the
scale $\Lambda_{i}$ by the formula%
\begin{equation}
D_{i}(m_{h})\equiv\left\vert \frac{\partial\log m_{h}^{2}}{\partial\log
\Lambda_{i}^{2}}\right\vert \,=\frac{|\alpha_{i}|\Lambda_{i}^{2}}{m_{h}^{2}}.
\label{Di}%
\end{equation}
The meaning of this quantity is that if $D_{i}>1$, the theory needs
fine-tuning of 1 part in $D_{i}$. The no fine-tuning condition $D_{i}\approx1$
is equivalent to demanding that quadratic contributions in (\ref{quad}) (taken
separately) do not exceed the physical mass squared. Using precise values of
$\alpha_{i}$ given in (\ref{quad}), we obtain three no fine-tuning scales (for
$D_{i}>1$ these scales should be multiplied by $\sqrt{D_{i}}$):%
\begin{align}
\Lambda_{t}  &  \approx3.5 \, m_{h}\nonumber\\
\Lambda_{g}  &  \approx9 \, m_{h}>\Lambda_{t}\\
\Lambda_{h}  &  \approx1.3\text{ TeV} . \nonumber\label{Li}%
\end{align}
These equations are the quantitative outcome of the naturalness
analysis---they bound the expected scale of the divergence-cancelling physics.
Not surprisingly, the precise value of this scale crucially depends on the
assumed value of the Higgs mass. The prevalent assumption nowadays is that the
Higgs is light, with $m_{h}$ close to the $114$ GeV limit from the direct
searches, so that the low value of $\Lambda_{t}$ makes us reasonably sure that
at least the physics cancelling the virtual top divergence should be seen at
the LHC.

But what if the Higgs is heavy, say $m_{h}\gtrsim400$ GeV? The scale
$\Lambda_{t}$ is raised above 1.4 TeV (``improved naturalness''), and since
$\Lambda_{h}$ is also rather large, we can no longer be certain that the
physics cancelling these divergences will be observable at the LHC. While
$\Lambda_{i}$ only provide upper bounds on the scale of the cancellation
physics, in the absence of supersymmetry, given the LEP paradox, it is likely
that these bounds are saturated. What will the LHC see in this case? This is
the question we would like to address.

\subsection{Perturbativity, or how heavy is heavy?}

\label{SM.p}How high up in $m_{h}$ can one go? As the Higgs mass is increased
so the quartic scalar interaction becomes stronger, and the maximum scale at
which perturbation theory is useful, $\Lambda_{P}$, is decreased. Our aim is
to have a natural theory up to energies of 1.5 TeV, hence we must require that
$\Lambda_{P} >$ 1.5 TeV, placing an upper bound on the Higgs mass.
If this requirement is fulfilled, we can reasonably assume that the
divergence-cancelling physics, which is expected to appear just around that
scale
will also be able to stop the growth of the Higgs quartic coupling and prevent
the Landau pole from appearing. The RG evolution of the Higgs quartic coupling
is reviewed in Appendix \ref{SM.pert}. The results of that discussion can be
summarized in terms of two scales: the one-loop Landau pole scale $\Lambda
_{L}$, and the perturbativity scale $\Lambda_{P}$ at which the quartic
coupling grows by 30\% from its value in the IR. The values of these two
scales for $m_{h}=400,500,600$ GeV are given in Table 1. We see that in all
cases $\Lambda_{P}$ is above 1.5 TeV, while $\Lambda_{L}$ is $5-30$ times
higher. The conclusion is that all masses in the $400-600$ GeV range are
suitable for the implementation of the improved naturalness idea.
\begin{table}[h]
\begin{center}%
\begin{tabular}
[c]{l|l|l}%
$m_{h},$GeV & $\Lambda_{P},$TeV & $\Lambda_{L}$,TeV\\\hline
400 & 2.4 & 80\\
500 & 1.8 & 16\\
600 & 1.6 & 7.5
\end{tabular}
\end{center}
\caption{{}Heavy Higgs perturbativity scale $\Lambda_{P}$ and Landau pole
$\Lambda_{L}$.}%
\end{table}

\subsection{ElectroWeak Precision Tests}

\label{SM.EWPT} 
\begin{figure}[ptb]
\begin{center}
\includegraphics[width=8cm]{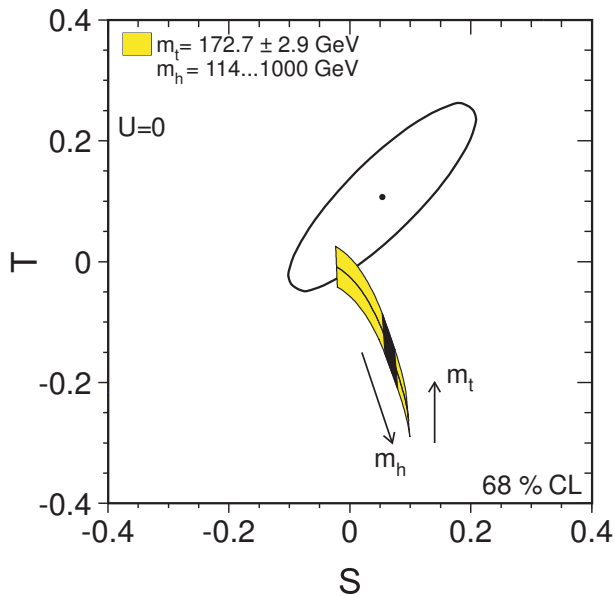}
\end{center}
\caption{(Adapted from \cite{Plot}.) Dependence of the $S,T$ parameters on the
Higgs mass. The thick black band marks $m_{h}=400-600$ GeV.}%
\label{STU}%
\end{figure}

At this point the reader should ask: but what about the EWPT,
which \textit{predict} that the Higgs is light? The answer of course is that
this `prediction' is true only in the absence of new physics, which may
contribute to the EWPT observables, but has nothing to do with cancelling the
quadratic divergences of the Higgs mass. Indeed, the Higgs mass influences the
EWPT via the logarithmic contributions to $T$ and $S$:%
\begin{align}
T  &  \approx-\frac{3}{8\pi c^{2}}\ln\frac{m_{h}}{m_{Z}}\\
S  &  \approx\frac{1}{6\pi}\ln\frac{m_{h}}{m_{Z}}. \label{ST}%
\end{align}
For large $m_{h}$ these contributions violate experimental constraints (see
Fig. \ref{STU}). Assuming that no new physics influences the EWPT, one obtains
$m_{h}=91_{-32}^{+45}$ GeV, with the upper bound $m_{h}<186$ GeV at 95\% CL
\cite{LEPEWWG}. In particular $m_{h}=400$ GeV is excluded at 99.9\% CL.

However, looking at Fig. \ref{STU} one immediately sees that the heavy Higgs
\textit{can} be consistent with the EWPT if there is new physics producing a
compensating positive $\Delta T$. If at the same time the $\Delta S$
contribution of this new physics is not too large, a good fit could be
obtained. For $m_{h}=400-600$ GeV (black band in Fig. \ref{STU}) the needed
compensating $\Delta T$ is
\begin{equation}
\Delta T\approx0.25\pm0.1, \label{need}%
\end{equation}
which would bring us near the central point of the 68\% CL ellipse (the
uncertainty in this number is mostly due to the experimental error on $T$).
Rather than making a careful fit, in this paper we will be content with this
rough estimate. 

Thus the answer to the question of what the LHC will see is: If the Higgs is
heavy, there must be new physics producing a positive $\Delta T$, and it is
this new physics that the LHC will study.

\section{The Inert Doublet Model}

\label{our}

In this section we will present what seems to us the most attractive
realization of the improved naturalness idea. Some alternatives are described
in Section \ref{alt}.

\subsection{The Model}

\label{our.1}

We consider the most general two-Higgs doublet model that possesses the
parity
\begin{equation}
H_{2} \rightarrow- H_{2} \label{parity}%
\end{equation}
with all other fields invariant. This parity imposes natural flavor
conservation in the Higgs sector\cite{GW}\footnote{In standard nomenclature
this would be called Type I 2HDM, except that we reverse the usual roles of
$H_{1}$ and $H_{2}$.}, implying that only $H_{1}$ couples to matter. The
scalar potential is
\begin{align}
V  &  =\mu_{1}^{2}|H_{1}|^{2}+\mu_{2}^{2}|H_{2}|^{2}+\lambda_{1}|H_{1}%
|^{4}+\lambda_{2}|H_{2}|^{4}+\lambda_{3}|H_{1}|^{2}|H_{2}|^{2}\nonumber\\
&  \quad+\lambda_{4}|H_{1}^{\dagger}H_{2}|^{2}+\frac{\lambda_{5}}{2}%
[(H_{1}^{\dagger}H_{2})^{2}+\text{h.c.}]. \label{pot}%
\end{align}
We assume that the parameters of this potential yield an asymmetric phase:
$H_{1}$ acquires a vev but $H_{2}$ does not\footnote{This phase of the
unbroken parity was considered recently in \cite{Ma} motivated by neutrino
physics. We thank E.~Ma for bringing this to our attention.} This is not the
well-studied standard phase of the theory that has both vevs non-zero, and it
cannot be obtained as the small $\tan\beta$ limit of the standard phase, which
is a fine tuned limit. Rather, the asymmetric phase results in a parameter
region of comparable size to the standard phase, depending essentially on the
sign of $\mu_{2}^{2}$. The doublet $H_{1}$ is identified as essentially the SM
Higgs doublet---it gets a vev and gives masses to $W,Z$ and fermions. On the
other hand, $H_{2}$ \textit{does not} couple to fermions and \textit{does not}
get a vev. We will call it the inert doublet, although of course it does have
weak interactions and quartic interactions.

The scalar spectrum of the theory is obtained by expanding the potential
around the minimum
\begin{equation}
H_{1}=(0,v),\quad H_{2}=(0,0). \label{min1}%
\end{equation}
The physical fields appear in the parametrizaton of the doublets as follows:%
\begin{equation}
H_{1}=\left(
\begin{array}
[c]{c}%
\phi^{+}\\
v+(h+i\chi)/\sqrt{2}%
\end{array}
\right)  ,\qquad H_{2}=\left(
\begin{array}
[c]{c}%
H^{+}\\
(S+iA)/\sqrt{2}%
\end{array}
\right)  . \label{param}%
\end{equation}
Here the Goldstones $\phi^{+},\chi$ can be put to zero by choosing the unitary
gauge; they are included for future reference. The usual Higgs boson is $h$,
which we take to be heavy:
\begin{equation}
m_{h}\approx400-600~\text{GeV\quad(}\lambda_{1}=m_{h}^{2}/4v^{2}\approx2).
\label{assumed}%
\end{equation}
In addition, we have three \textquotedblleft inert\textquotedblright%
\ particles---a charged scalar $H^{+}$ and two neutrals $S,A$ with masses:%
\begin{align}
m_{I}^{2}  &  =\mu_{2}^{2}+\lambda_{I}v^{2},\quad I=\{H,S,A\}\label{masses}\\
\lambda_{H}  &  =\lambda_{3}\nonumber\\
\lambda_{S}  &  =\lambda_{3}+\lambda_{4}+\lambda_{5}\nonumber\\
\lambda_{A}  &  =\lambda_{3}+\lambda_{4}-\lambda_{5}.\nonumber
\end{align}
We assume that the potential (\ref{pot}) is bounded from below, which happens
if and only if%
\begin{equation}
\lambda_{1,2}>0;\qquad\lambda_{3},\lambda_{\text{L}}\equiv\lambda_{3}%
+\lambda_{4}-|\lambda_{5}|>-2(\lambda_{1}\lambda_{2})^{1/2}. \label{stab}%
\end{equation}
Under this assumption, the minimum (\ref{min1}) is stable and global, as long
as all masses squared (\ref{masses}) are positive.

The way to visualize the parameter space of the 7 parameters of the potential
(\ref{pot}) is as follows. These 7 parameters can be traded for the four
physical scalar masses, $m_{h},m_{H},m_{A},m_{S}$, the vev $v$ (or the Z-mass)
and the two quartic couplings, $\lambda_{2}$ and $\lambda_{3}$. The EWPT imply
a relation between the 5 parameters with dimension of mass, analogous to the
relation between $m_{h}$ and $m_{Z}$ in the SM. Since the inert parity,
(\ref{parity}), is unbroken, the lightest inert particle (LIP) will be stable
and will contribute to the Dark Matter density. It may in fact constitute all
of the DM if the parameters have the right value, although the typical
fraction is small. In any case, to avoid conflicting with the stringent limits
on charged relics \cite{charged}, we will always assume that \textit{the LIP
is neutral}\footnote{This can be avoided only by considering the parity
(\ref{parity}) to be an approximate symmetry.}. In the limit of Peccei-Quinn
symmetry, $\lambda_{5}\rightarrow0$, the neutral inert scalars $S$ and $A$
become degenerate. Direct detection of halo dark matter places a limit on this
degeneracy \cite{CDMS}, because the mass difference must be sufficient to
kinematically suppress the scattering of galactic LIPs on nuclei via
tree-level Z boson exchange.


Of the two dimensionless couplings, $\lambda_{2}$ only affects the
self-interactions between the inert particles. It is difficult to even
conceive how it could be measured. To avoid additional problems with
perturbativity, we assume that it is quite small,
\begin{equation}
\lambda_{2}\lesssim1\,. \label{l2}%
\end{equation}
On the contrary, $\lambda_{3}$ may affect some significant observables, like
the width of $h$ (see Eq. (\ref{incr})) and (if parameters take values to
allow LIP DM) the interaction cross section of the DM with nuclei (see Eq.
(\ref{higgsexch})).

Analogously to the SM case, in the next sub-sections we discuss constraints
imposed on the IDM parameters by perturbativity, naturalness, and the EWPT,
and we summarize the allowed regions of couplings in Section \ref{summary}. In
a large region of parameter space we will find that the heavy Higgs has
naturalness and perturbativity properties very similar to the SM heavy Higgs
described in section 2. The advantage of the IDM is that the mass splittings
within the inert doublet allow a satisfactory $T$ parameter.

\subsection{Perturbativity}

\label{our.pert}Let us begin with perturbativity. The RG equations satisfied
by the two-Higgs doublet model couplings are given in Appendix \ref{2HDM.rg}.
To determine the exact high-energy behavior, one would have to find precise
initial conditions for all couplings, similarly to what we have done for the
SM in Appendix \ref{SM.pert}. Here we will be content with deriving some
sufficient conditions for perturbativity. First let us look at $\lambda_{1}$,
whose beta-function equation is%
\begin{equation}
16\pi^{2}\frac{d\lambda_{1}}{d\log\Lambda}=24\lambda_{1}^{2}+2\lambda_{3}%
^{2}+2\lambda_{3}\lambda_{4}+\lambda_{4}^{2}+\lambda_{5}^{2}. \label{rg1}%
\end{equation}
As we discussed in Section \ref{SM.p}, the SM with a 500 GeV Higgs stays
perturbative up to a reasonably high scale $\sim1.8$ TeV. In order that this
conclusion be preserved in our model, we will impose a requirement that the
sum the of extra terms in the RHS of (\ref{rg1}) not exceed 50\% of the
SM\ term $24\lambda_{1}^{2}$. Thus we get a constraint (see Eq. (\ref{assumed}%
))%
\begin{equation}
|2\lambda_{3}(\lambda_{3}+\lambda_{4})+\lambda_{4}^{2}+\lambda_{5}%
^{2}|\lesssim50\text{ \ \ \ (perturbativity).} \label{cp}%
\end{equation}
How large can the couplings become consistent with this inequality? One
possibility is that $|\lambda_{4}|$ becomes large, while $\lambda_{5}$ stays
relatively small. In this case we must have $\lambda_{4}<0$ for the LIP to be
neutral. This implies that $\lambda_{3}$ must also become large, $\lambda
_{3}\gtrsim|\lambda_{4}|$, to ensure the vacuum stability (\ref{stab})
(remember that $\lambda_{2}$ is assumed to be small.) The critical region is
when $\lambda_{3}\sim|\lambda_{4}|$ so that the first term in (\ref{cp})
vanishes. This way we get the bound
\begin{equation}
|\lambda_{4}|\lesssim\lambda_{3}\lesssim7\quad(\lambda_{4}<0\text{, }%
\lambda_{5}\text{ small).} \label{pert1}%
\end{equation}

The other possibility is that, on the contrary, it is $\lambda_{5}$ which
becomes large, while $\lambda_{4}$ is small. In this case, the vacuum
stability condition (\ref{stab}) implies $\lambda_{3}\gtrsim|\lambda_{5}|$,
leading to a stricter bound
\begin{equation}
|\lambda_{5}|\lesssim\lambda_{3}\lesssim4\text{\qquad(}\lambda_{4}\text{
small).} \label{pert2}%
\end{equation}

As we will see in Section \ref{our.EWPT} below, it is only the first
possibility that will lead to $\Delta T>0$, as needed to compensate for the
heavy Higgs. However, for the time being we want to explore all possibilities
so that we can understand the typical range of $\Delta T$ allowed in our model.

Finally, we have checked that, in the region allowed by the constraints
(\ref{pert1}) or (\ref{pert2}), the evolution of the remaining couplings does
not lead to any additional restrictions. Essentially this happens because we
assume that $\lambda_{2}$ is sufficiently small and because $\lambda_{3,4,5}$
evolve slower than $\lambda_{1}$ due to the smaller RG
coefficients\footnote{Also, in the first case, $\lambda_{3}$ grows faster than
$|\lambda_{4}|$ in the UV, and thus the vacuum stability is preserved.}.

\subsection{Naturalness}

\label{our.natur}

Like the SM, the IDM is a natural effective theory only up to some cutoff,
which is determined by the quadratic divergences in the dimensional
parameters. For the IDM there are two mass parameters, $\mu_{1,2}^{2}$, and we
must study naturalness for each separately, obtaining conditions that allow
the theory to be natural for energies up to 1.5 TeV.

Since $\mu_{1}^{2}$ is linear in the Higgs mass squared, as in the SM case it
is convenient to study the corrections to $m_{h}^{2}$. Introducing separate
cutoffs for loops of virtual $H_{1}$ and $H_{2}$ particles, $\Lambda_{H_{1,2}%
}$, we find a result similar to (\ref{quad})
\begin{equation}
\delta m_{h}^{2}=\alpha_{t}\Lambda_{t}^{2}+\alpha_{g}\Lambda_{g}^{2}%
+\alpha_{11}\Lambda_{H_{1}}^{2}+\alpha_{12}\Lambda_{H_{2}}^{2} \label{quad2}%
\end{equation}
where
\begin{equation}
\alpha_{11}=-\frac{3\lambda_{1}}{4\pi^{2}},\;\;\alpha_{12}=-\frac{2\lambda
_{3}+\lambda_{4}}{8\pi^{2}} \label{alpha11}%
\end{equation}
and $\alpha_{t,g}$ are as in (\ref{alphat}). The first three terms lead to the
bounds of (\ref{Li}), except that it is now $\Lambda_{H_{1}}$, rather than
$\Lambda_{h}$, that is limited by 1.3 TeV. This last bound cannot be avoided
without changing or cancelling the effect of the usual Higgs quartic, which
the IDM does not do. This is why we content ourselves with a theory that is
natural up to about 1.5 TeV. The scales $\Lambda_{t,g}$ are raised to 1.5 TeV
or more by taking the Higgs mass heavier than 400 GeV. Requiring that the last
term of (\ref{quad2}) not exceed the physical Higgs mass squared gives the
additional constraint
\begin{equation}
|2\lambda_{3}+\lambda_{4}|\lesssim9. \label{nat1}%
\end{equation}

The one-loop quadratic divergences to $\mu_{2}^{2}$ are
\begin{equation}
\delta\mu_{2}^{2} = - \frac{1}{2} \left(  \alpha_{g} \Lambda_{g}^{2}%
+\alpha_{22} \Lambda_{H_{2}}^{2} + \alpha_{21} \Lambda_{H_{1}}^{2} \right)
\label{quadmu2}%
\end{equation}
where
\begin{equation}
\alpha_{22} = -\frac{3 \lambda_{2}}{4 \pi^{2}}, \; \; \; \; \alpha_{21} =
-\frac{2 \lambda_{3} + \lambda_{4}}{8 \pi^{2}}. \label{alpha22}%
\end{equation}
Requiring each of these three corrections to be smaller than the tree-level
value, leads to the three naturalness constraints
\begin{equation}
\mu_{2} \gtrsim\left(  1,\; 2.5 \sqrt{\lambda_{2}}, \; \sqrt{|2\lambda_{3} +
\lambda_{4}|} \right)  120 \, \mbox{GeV} \label{mu2nat}%
\end{equation}
respectively.

We have required that our model is a natural effective field theory in the
sense that the sensitivity of Lagrangian parameters to variations in the
cutoff is small: $D_{i}(\mu_{1,2}^{2})\lesssim1$. We do not attempt to impose
the stronger condition that all observables have small such sensitivities. It
may be that some observables are small because of cancelling contributions
within the effective theory. For example, from (\ref{masses}) we see that a
LIP mass $m_{\text{L}}\ll\mu_{2}$ requires a cancellation between $\mu_{2}%
^{2}$ and $\lambda_{3,4,5}v^{2}$ terms. Another example is the Z boson mass in
the minimal supersymmetric standard model with a heavy top squark. While these
cancellations should also be avoided, they differ from the cancellations at
the cutoff that are required between tree and loop contributions to Lagrangian
parameters. In particular they become acceptable if it is possible to measure
sufficient quantities to demonstrate that such cancellations occur in the low
energy theory. Given the expression (\ref{masses}) for the inert scalar
masses, it is natural to expect that some inert scalars could be somewhat
lighter than $\mu_{2}$, and some could be heavier. Since it is reasonable that
the terms in (\ref{masses}) for the LIP do not all have the same sign, it is
certainly natural for the LIP to be lighter than $\mu_{2}$. We consider
$m_{\text{L}}$ to be natural if
\begin{equation}
m_{\text{L}}\gtrsim\frac{\mu_{2}}{2}. \label{mL}%
\end{equation}

\subsection{ElectroWeak Precision Tests}

\label{our.EWPT}Finally, let us evaluate the IDM from the EWPT viewpoint. The
heavy Higgs contributions to $T$ is given in (\ref{ST}) and is to be
compensated by the contribution from the inert doublet, which is computed in
Appendix \ref{App.EWPT} to be
\begin{align}
\Delta T  &  =\frac{1}{32\pi^{2}\alpha v^{2}}[F(m_{H},m_{A})+F(m_{H,}%
m_{S})-F(m_{A},m_{S})],\label{extra}\\
&  F(m_{1},m_{2})=\frac{m_{1}^{2}+m_{2}^{2}}{2}-\frac{m_{1}^{2}m_{2}^{2}%
}{m_{1}^{2}-m_{2}^{2}}\ln\frac{m_{1}^{2}}{m_{2}^{2}}.
\end{align}
This contribution comes from the $\lambda_{4,5}$ terms in the potential, since
these are the terms breaking the custodial symmetry. From (\ref{masses}), it
is clear that the same terms are responsible for the mass splitting among the
inert scalars. The function $F(m_{1},m_{2})$ is positive, symmetric, vanishes
for $m_{1}=m_{2}$ and monotonically increases for $m_{1}\geq m_{2}$. Moreover,
to high accuracy, 2$\ldots$5\% for $1\leq m_{1}/m_{2}\leq2\ldots3$, we have
\begin{equation}
F(m_{1},m_{2})\approx\frac{2}{3}(m_{1}-m_{2})^{2}.
\end{equation}
For our purposes it will always be sufficient to use this approximation,
allowing (\ref{extra}) to be simplified
\begin{equation}
\Delta T\approx\frac{1}{24\pi^{2}\alpha v^{2}}(m_{H}-m_{A})(m_{H}-m_{S}).
\label{dT1}%
\end{equation}
Requiring this $\Delta T$ be in the range (\ref{need}), we find a constraint
on the spectrum
\begin{equation}
(m_{H}-m_{S})(m_{H}-m_{A})=M^{2},\qquad M=120_{-30}^{+20}\text{ GeV}\,.
\label{pred}%
\end{equation}
Since the LIP is neutral, we see that $H$ should be heavier than both $S$ and
$A$ to have $\Delta T>0$.

The contribution of the inert doublet to $S$ is also given in Appendix
\ref{App.EWPT}, Eq. (\ref{dS}). It depends on the inert particle masses only
logarithmically, and remains small ($|\Delta S|\lesssim0.04$) for the whole
range of parameters considered below$.$ Thus its effect on the EWPT fit can be neglected.

\begin{figure}[ptb]
\begin{center}
\includegraphics[width=8cm]{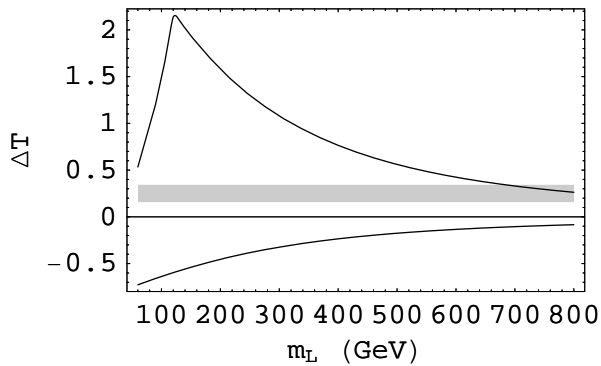}
\end{center}
\caption{{}The maximal and minimal $\Delta T$ allowed by naturalness and
perturbativity as a function of $m_{L}$. The horizontal grey band marks the
range needed to compensate for the heavy Higgs.}%
\label{range}%
\end{figure}

To evaluate the success of the IDM in compensating for the heavy Higgs, it is
important to know the typical range of $\Delta T$ allowed by naturalness and
perturbativity. The relevant constraints on the parameters are (\ref{stab}%
,\ref{l2},\ref{pert1},\ref{pert2},\ref{mu2nat},\ref{mL}). The resulting
$\Delta T$ range is shown as a function of $m_{\text{L}}$ in Fig.~\ref{range}.
For $m_{\text{L}}\gtrsim300$ GeV the perturbativity constraints (\ref{pert1}%
,\ref{pert2}) are more restrictive, while for smaller $m_{\text{L}}$ the
naturalness constraints become crucial. The maximal $\Delta T>0$ occurs when
$\lambda_{4}$ is large and negative, while $\lambda_{5}$ remains small. The
maximal $\Delta T<0$ is achieved in the opposite regime of $\lambda_{4}$
small, $\lambda_{5}$ large. We see that $\Delta T$ is predominantly positive
and is of the typical size needed to compensate for the heavy Higgs in a large
region of the parameter space. We conclude that the success of our model is
\textit{not} accidental. If it had turned out that the needed $\Delta T$ was
much smaller than the typical value, then we would have imposed approximate
custodial symmetry on the potential. But we see that little, if any,
suppression from custodial symmetry is needed in most of the range of
$m_{\text{L}}$\footnote{If we, say, insist on a stricter upper bound
$|\lambda_{4}|\lesssim2$, then $\Delta T_{\max}$ is lowered to $0.6.$}.

\subsection{Summary of constraints on the spectrum and couplings}

\label{summary} 
\begin{figure}[ptb]
\begin{center}
\includegraphics[width=8cm]{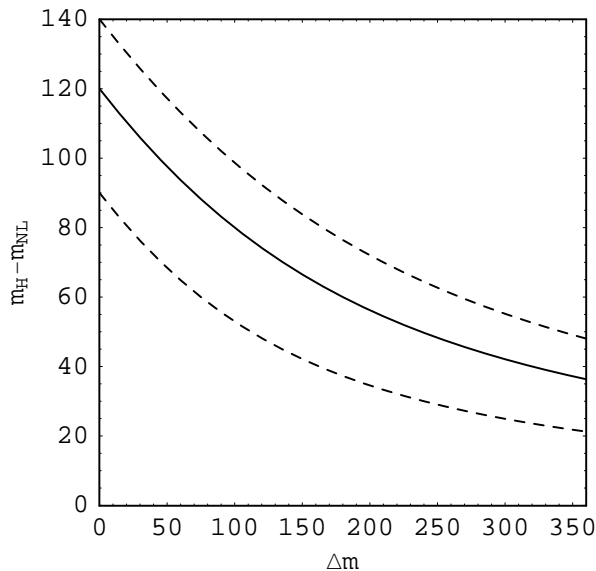}
\end{center}
\caption{The relation between the first and the second spacing in the inert
particle spectrum following from the EWPT constraint (\ref{pred}). }%
\label{spacings}%
\end{figure}

Preparing for the discussion of signals, let us describe the
region of parameter space that leads to a natural, perturbative effective
theory up to 1.5 TeV and that satisfies the EWPT constraint (\ref{pred}). It
is convenient to use a parametrization in terms of the masses of the two
neutral inert particles, $m_{\text{L}}$ for the lightest and $m_{\text{NL}}$
for the next-to-lightest. We consider the general case when $\Delta
m=m_{\text{NL}}-m_{\text{L}}$ can be sizeable. The charged scalar is always
heavier than both neutrals, and using (\ref{pred}), the second splitting can
be expressed in terms of $\Delta m$ and $M$
\begin{equation}
m_{H}-m_{\text{NL}}=\sqrt{M^{2}+\frac{(\Delta m)^{2}}{4}}-\frac{\Delta m}{2}
\label{*}%
\end{equation}
and is shown in Fig.~\ref{spacings} for the range $M= 120^{+20}_{-30}$ GeV.

The couplings $\lambda_{4,5}$ can be also expressed via $m_{\text{L}%
},m_{\text{NL}}$ using (\ref{*}) and (\ref{masses}), giving%
\begin{align}
\lambda_{4}  &  =-\frac{1}{v^{2}}\left(  M^{2}+(m_{\text{L}}+m_{\text{NL}%
})\sqrt{M^{2}+\frac{(\Delta m)^{2}}{4}}\right)  <0\label{l4}\\
|\lambda_{5}|  &  =\frac{m_{\text{NL}}^{2}-m_{\text{L}}^{2}}{2v^{2}}%
<|\lambda_{4}|. \label{l5}%
\end{align}
The sign of $\lambda_{5}$ depends on whether it is the scalar $S$ or the
pseudoscalar $A$ which is the heavier.

\begin{figure}[ptb]
\begin{minipage}[c]{0.5\linewidth}
\centering
\includegraphics[width=8cm]{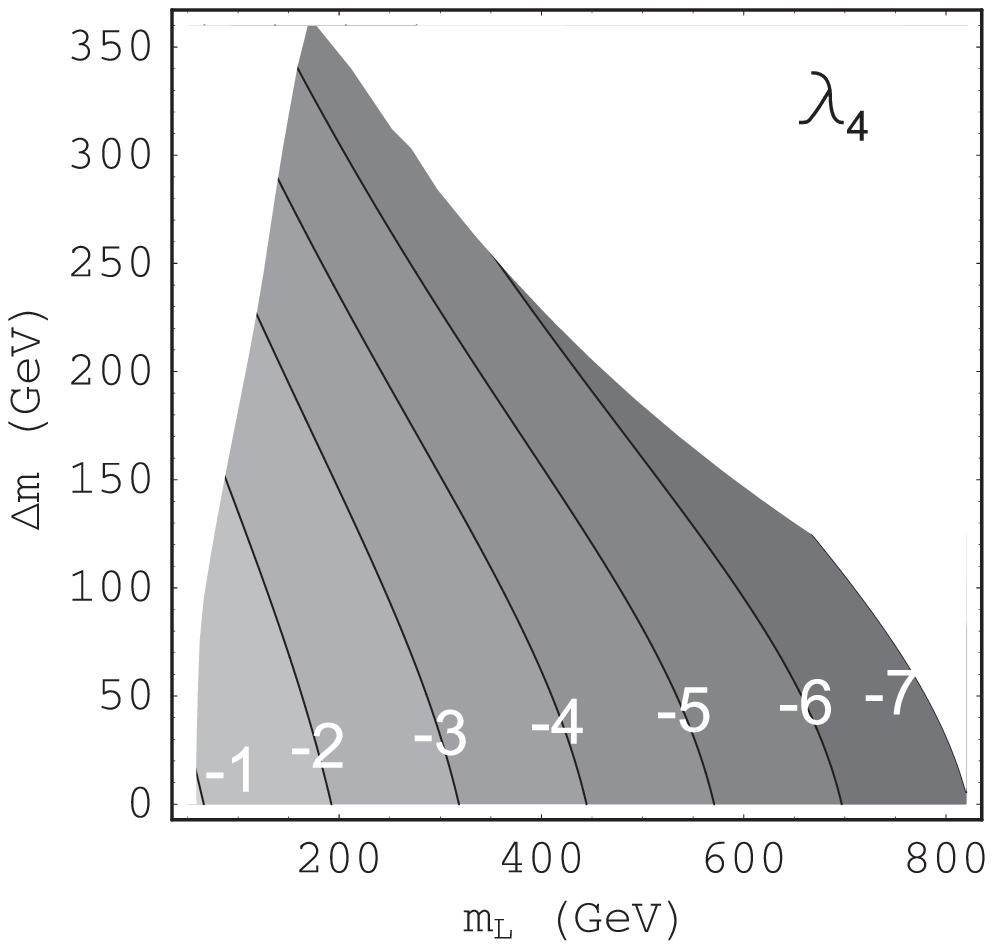}
\end{minipage}
\begin{minipage}[c]{0.5\linewidth}
\centering
\includegraphics[width=8cm]{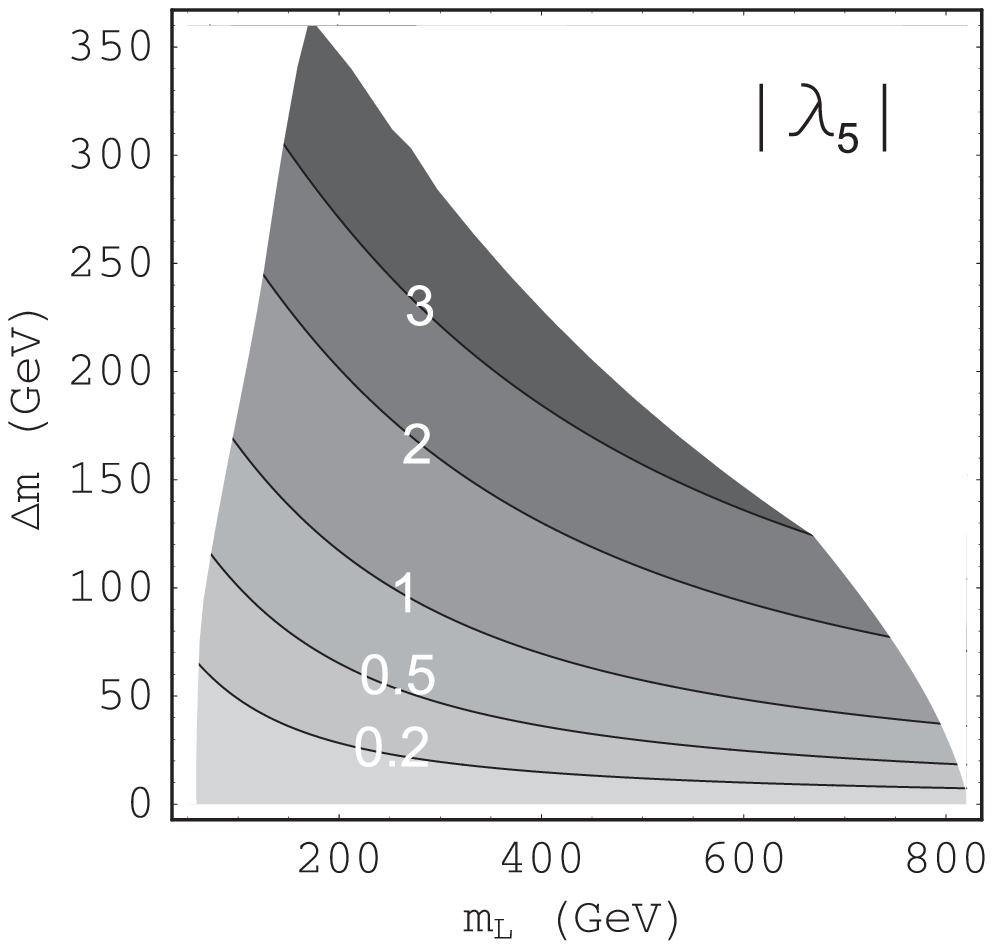}
\end{minipage}
\caption{ $\lambda_{4,5}$ in the allowed region as functions of $m,\Delta m$
computed from (\ref{l4}), (\ref{l5}) with $M=120$ GeV. }%
\label{lambda45}%
\end{figure}

\begin{figure}[ptb]
\begin{minipage}[c]{0.5\linewidth}
\centering
\includegraphics[width=8cm]{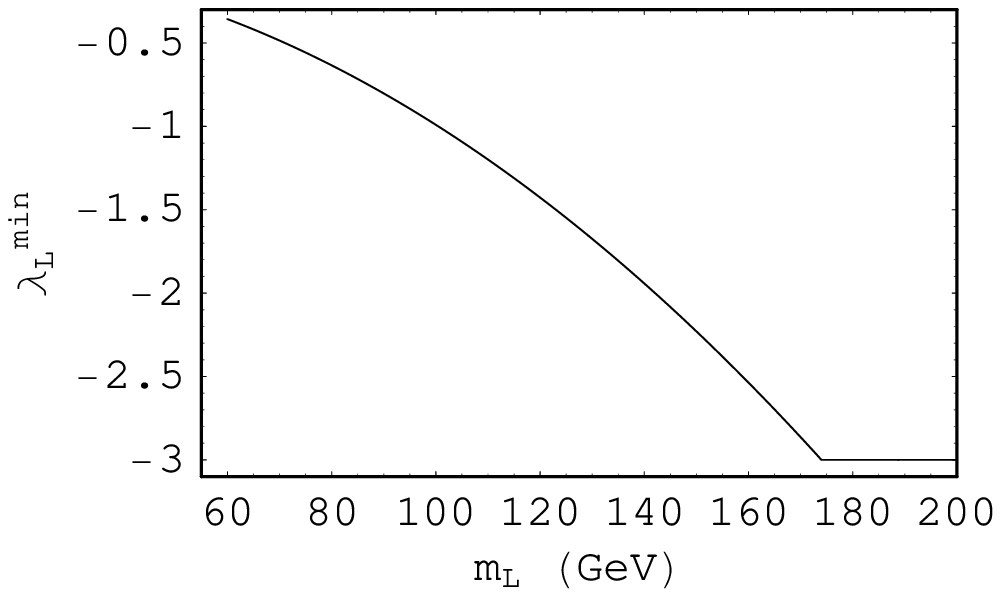}
\end{minipage}
\begin{minipage}[c]{0.5\linewidth}
\centering
\includegraphics[width=8cm]{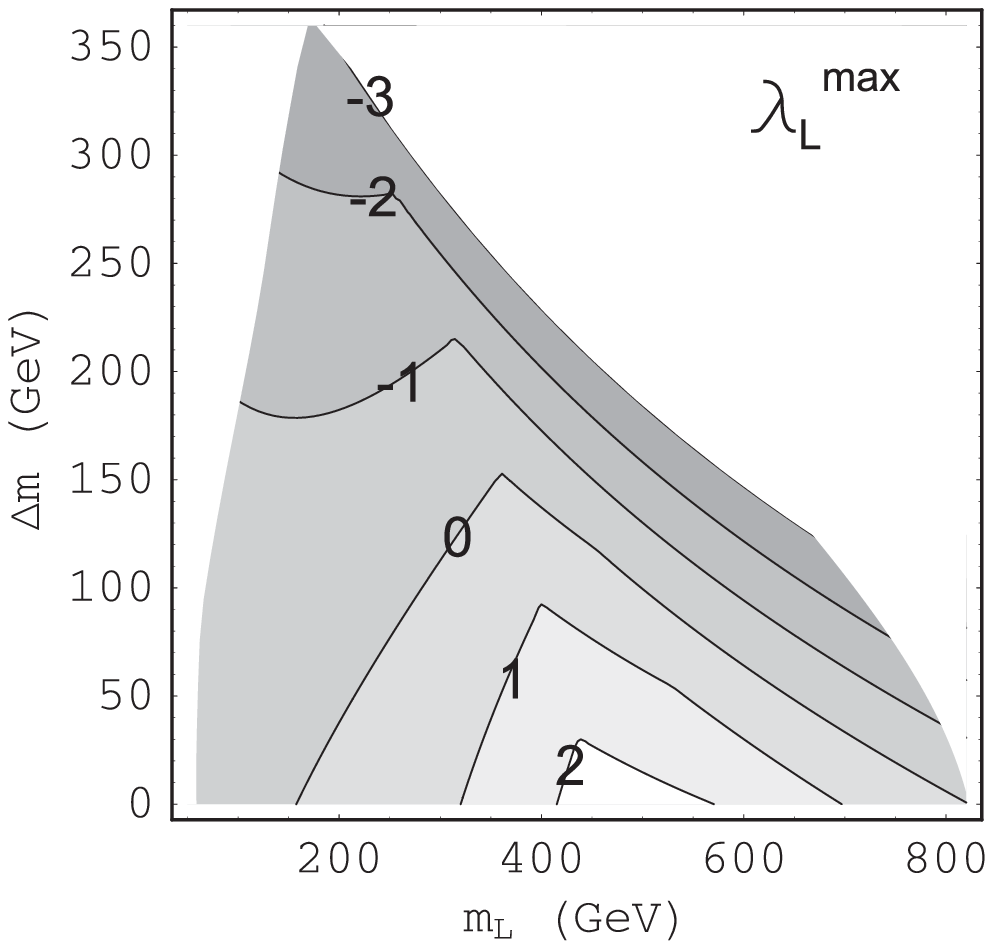}
\end{minipage}
\caption{ $\lambda_{\text{L}}^{\min}$ and $\lambda_{\text{L}}^{\max}$ \ for
$M=120$ GeV. Notice that $\lambda_{\text{L}}^{\min}$ depends only on
$m_{\text{L}}$ and is constant for $m_{\text{L}}\gtrsim180$ GeV. }%
\label{Dminmax}%
\end{figure}

The coupling $\lambda_{3}$ (or $\lambda_{\text{L}}\equiv\lambda_{3}%
+\lambda_{4}-|\lambda_{5}|$) is the only free parameter; it should be chosen
in agreement with the perturbativity (\ref{l2},\ref{pert1}), naturalness
(\ref{nat1},\ref{mu2nat},\ref{mL}), and vacuum stability (\ref{stab})
constraints. These constraints can be used to derive a range of allowed
values:%
\begin{equation}
\lambda_{\text{L}}^{\min}(m_{\text{L}})\lesssim\lambda_{\text{L}}%
\lesssim\lambda_{\text{L}}^{\max}(m_{\text{L}},\Delta m) \label{range1}%
\end{equation}
The (iso)plots of $\lambda_{4,5}$ $\lambda_{\text{L}}^{\min},\lambda
_{\text{L}}^{\max}$ are given in Fig.~\ref{lambda45},\ref{Dminmax}. The white
region has $\lambda_{\text{L}}^{\min}>\lambda_{\text{L}}^{\max}$ or
$|\lambda_{4}|>7$ and is disfavored by naturalness and/or perturbativity.

Thus we conclude that the IDM is a fully natural effective field theory up to
1.5 TeV for a large region of parameter space where the Higgs is heavy and
EWPT are satisfied. Of the 7 parameters in the potential, $\mu_{1}^{2}$ and
$\lambda_{1}$ can be traded for $v$ and $m_{h}$, while $\mu_{2}^{2}$ and
$\lambda_{3,4,5}$ can be traded for $m_{\text{L}},m_{\text{NL}},m_{H},$ and
$\lambda_{\text{L}}$. EWPT constrains $m_{\text{H}}-m_{\text{NL}}$ as shown in
Figure 3. The allowed ranges of $m_{\text{L}}$ and $\Delta m$ are shown shaded
in Figure 4, and the allowed range of $\lambda_{\text{L}}$ is shown in Figure
5. From perturbativity, $\lambda_{2} \lesssim1$.

\subsection{Dark Matter}

\label{our.DM}

We now begin the discussion of signals. As we already mentioned, the LIP is
stable, and thus provides a cold Dark Matter candidate\footnote{A possibility
also mentioned in \cite{Ma}.}. Here we will estimate its relic abundance and
discuss prospects for direct detection.

\subsubsection{Relic abundance}

\textbf{Case I:} $m_{\text{L}}\gtrsim m_{W}$

This case is of significant interest, since it includes most of the region of
parameter space preferred by naturalness. The dominant annihilation process is
into gauge bosons, with \textit{s}-wave cross section $\sigma_{\text{ann}%
}v_{\text{rel}}\sim130$ pb for $m_{\text{L}}\sim m_{W}$, decreasing to
$\sim10$ pb for $m_{\text{L}}\sim400$ GeV (see Appendix \ref{DM.1}). A
particular feature of our model is that the cross section does not decrease
further due to the contribution of the longitudinal final states. Using the
standard formalism \cite{KT}, we find the relic density $\Omega_{\text{DM}%
}h^{2}\lesssim0.02$ in the whole range of $m_{\text{L}}$, decreasing to
$0.002$ for $m_{\text{L}}\sim m_{W}$. This number can be trusted as an order
of magnitude estimate all the way down to the $WW$ production threshold. Since
this is much lower than the observed $\Omega_{\text{DM}}h^{2}\sim0.1$, we
conclude that in this region of parameter space the LIP provides only a
sub-dominant component of the Dark Matter.

\textbf{Case II:} $m_{\text{L}}<m_{W}$

Let us focus on the region $m_{\text{L}}= (60 \text{--} 80)$ GeV. While some
cancellations in (\ref{masses}) for the LIP mass are required, they are mild
and satisfy (\ref{mL}).

As the temperature of the early universe falls well below $m_{L}$, thermal
equilibrium is maintained via \textit{p}-wave suppressed coannihilations of
$S$ and $A$ into fermions, and the relic abundance critically depends on
$\Delta m$. In appendix \ref{DM.2}, we find the thermally averaged cross
section, for $\Delta m\ll T_{\text{f}}\sim m/25$, to be $\langle
\sigma_{\text{coann}}v_{\text{rel}}\rangle\sim(60 \text{--} 15)$ pb, for
$m_{\text{L}}= (60 \text{--} 80)$ GeV. This leads to the relic density
$\Omega_{\text{DM}}h^{2}\approx(0.5 \text{--} 2.5) \times10^{-2}$, still below
the observed value. On the other hand, for $\Delta m\gtrsim T_{\text{f}}$ the
density of the heavier component is thermally suppressed and the
coannihilation rate decreases. A formalism to compute the relic abundance in
such non-standard situations was developed in \cite{Griest}. However, the
final result can be predicted without making difficult calculations. Roughly,
the resulting relic density will be a factor $\sim(1/2)\exp(\Delta
m/T_{\text{f}})$ larger than in the unsplit case, where $1/2$ takes into
account that only the lighter component now contributes to the final
abundancy. This way we deduce that $\Delta m_{\text{naive}}\approx8$ GeV
should be enough to yield the observed DM density.

\begin{figure}[ptb]
\centering
\includegraphics[width=8cm]{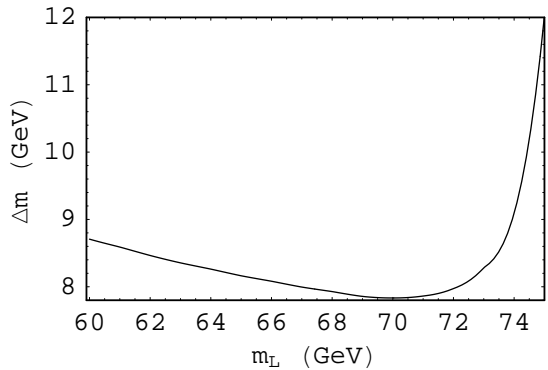} \caption{ The mass splitting between the
neutral inert particles needed to get the observed DM abundance below the WW
threshold (obtained using the formalism from \cite{Griest}). For $m_{\text{L}%
}\geq76$ GeV no splitting works.}%
\label{DMdiff}%
\end{figure}

The above naive argument can be expected to work at least for $m_{\text{L}%
}\lesssim m_{W}-3T_{\text{f}}/2\approx75$ GeV; for higher masses the
annihilation into $WW$ becomes thermally allowed and suppresses the relic
abundance. Using the above-mentioned formalism of \cite{Griest}, these numbers
can be confirmed (see Fig.~\ref{DMdiff}). In particular we find $\Delta
m\approx(8\mbox{--}9)$ GeV for $m_{\text{L}}=(60\mbox{--}73)$ GeV,
\ increasing to $12$ GeV for $m_{\text{L}}=75$ GeV, while for $m_{\text{L}%
}\geq76$ GeV no splitting gives the observed DM density.

\subsubsection{Direct detection}

The $S$ and $A$ have a vector-like interaction with the $Z$-boson, which
produces a spin-independent elastic cross section on a nucleus $\mathcal{N}$%
\begin{equation}
\sigma_{Z}(\text{L\thinspace}\mathcal{N}\rightarrow\text{L}\,\mathcal{N}%
)=\frac{G^{2}_{\text{F}}m_{r}^{2}}{2\pi}[N-(1-4s_{\text{w}}^{2})Z]^{2}\text{,}
\label{Zexch}%
\end{equation}
where $N$ and $Z$ are the numbers of neutrons and protons in the nucleus, and
$m_{r}=m_{L}m_{\mathcal{N}}/(m_{L}+m_{\mathcal{N}})$ is the reduced mass. The
resulting per nucleon cross section is 8-9 orders of magnitude above the
existing limits \cite{CDMS}. Thus we have to assume that there exists a
non-zero splitting between $S$ and $A$ larger than the kinetic energy of
DM\ in our galactic halo, so that the process (\ref{Zexch}) is forbidden
kinematically. This constraint must be imposed whether $m_{\text{L}}$ is above
or below $m_{W}$---even though the LIP relic density for $m_{\text{L}}\gtrsim
m_{W}$ is small, it is still too large to allow elastic scattering from nuclei
via tree-level $Z$-exchange.

Tree-level $h$ exchange produces a spin-independent cross section
\cite{Barbieri:1988zs}:%
\begin{equation}
\sigma_{h}(\text{L\thinspace}\mathcal{N}\rightarrow\text{L\thinspace
}\mathcal{N})=\frac{m_{r}^{2}}{4\pi}\left(  \frac{\lambda_{\text{L}}%
}{m_{\text{L}}m_{h}^{2}}\right)  ^{2}f^{2}m_{\mathcal{N}}^{2}\text{,}
\label{higgsexch}%
\end{equation}
where $f\sim0.3$ is the usual nucleonic matrix element:%
\begin{equation}
\langle\mathcal{N}|\sum m_{q}q\bar{q}|\mathcal{N}\rangle=fm_{\mathcal{N}%
}\langle\mathcal{N}|\mathcal{N}\rangle.
\end{equation}

Another allowed process, the exchange of two gauge bosons at one loop, gives
an effective coupling to nucleons similar to the tree-level $h$ exchange (see
\cite{MDM} for a recent discussion). For $m_{\text{NL}}-m_{\text{L}}\ll m_{Z}%
$, the resulting spin-independent cross section is independent of $m_{\text{L
}}$and can be estimated as%
\begin{equation}
\sigma_{\text{VV}}(\text{L\thinspace}\mathcal{N}\rightarrow\text{L\thinspace
}\mathcal{N})\sim\frac{m_{r}^{2}}{4\pi}\left[  \frac{(g/2c_{\text{w}})^{4}%
}{16\pi^{2}m_{Z}^{3}}\right]  ^{2}f^{2}m_{\mathcal{N}}^{2}, \label{VV}%
\end{equation}
while for larger splittings a cross section estimate can be obtained by
replacing $m_{Z}^{3}$ in the amplitude by $m_{\text{L}}m_{Z}^{2}$.

The numerical value of the cross-section (\ref{higgsexch}) for scattering from
a proton is
\begin{equation}
\sigma_{h}(\text{L}p\rightarrow\text{L}p)\approx2\times10^{-9}%
\;\mbox{pb}\;\left(  \frac{\lambda_{L}}{0.5}\right)  ^{2}\left(  \frac{70 \,
\mbox{GeV}}{m_{L}}\right)  ^{2}\left(  \frac{500 \, \mbox{GeV}}{m_{h}}\right)
^{4}. \label{higgsexchnum}%
\end{equation}
Our mass choices follow because the relic LIP abundance can yield the observed
DM for $m_{L}\approx$ 60--75 GeV, and the cutoff scales of the theory are
quite high if the Higgs is heavy, $m_{h}\approx$ 400--600 GeV. These ranges
for $m_{L}$ and $m_{h}$ do lot lead to a wide variation of $\sigma_{h}$. The
largest uncertainty in $\sigma_{h}$ arises from $|\lambda_{L}|=(\mu_{2}%
^{2}-m_{L}^{2})/v^{2}$. From (\ref{mu2nat}) and (\ref{mL}) naturalness
suggests that, for this interesting case of a light LIP, $\mu_{2}$ should be
close to its lowest natural value of 120 GeV, giving $|\lambda_{L}|\approx
0.5$, the value used in (\ref{higgsexchnum}). In this region of parameter
space, the cross section as estimated in (\ref{VV}) is typically an order of
magnitude smaller. Thus we expect a signal two orders of magnitude below the
present limit from Ge detectors \cite{CDMS} and within the sensitivity of
experiments currently under study.

Finally, for $m_{\text{L}}\gtrsim m_{W}$ we are penalized by a smaller relic
density and by the $m_{\text{L}}^{-2}$ decrease of (\ref{higgsexch}). The
prospects for near-future direct detection in this case are dim.

\subsection{Collider signals}

\subsubsection{Production and decay of the inert particles}

\label{coll}

The inert particles can be only pair-produced. If $m_{\text{L}}\approx70$ GeV
and $\Delta m$ is small, as preferred in the DM region, $SA$ pairs were
produced at LEP2. Assuming $\Delta m\ll m_{\text{L}}$, the production cross
section is%
\begin{equation}
\sigma(e^{+}e^{-}\rightarrow SA)=\left(  \frac{g}{2c_{\text{w}}}\right)
^{4}\left(  \frac{1}{2}-2s_{\text{w}}^{2}+4s_{\text{w}}^{4}\right)  \frac
{1}{48\pi s}\frac{[1-4m^{2}/s]^{3/2}}{[1-m_{Z}^{2}/s]^{2}}\approx0.2\text{ pb}
\label{prod_lep}%
\end{equation}
for $\sqrt{s}=200$ GeV. The heavier state, which for definiteness we take to
be $A$, decays into the lighter plus $Z^{\ast}$. The resulting dilepton events
with missing energy were looked for in the context of searches for the
lightest superpartner. For small mass differences, $\Delta m\lesssim10$ GeV,
the production cross section (\ref{prod_lep}) is below the existing limits set
by the separate LEP collaborations \cite{limits}. However, our signal is close
to these limits, so that a combined reanalysis of the old data may be useful.

At the LHC pairs of inert particles will be produced by%
\begin{align}
pp  &  \rightarrow W^{\ast}\rightarrow HA\text{ or }HS\label{HS}\\
pp  &  \rightarrow Z^{\ast}(\gamma^{\ast})\rightarrow SA\text{ or }H^{+}H^{-}%
\end{align}
and will decay by%
\begin{align}
H  &  \rightarrow AW\text{ or }SW\label{Hd}\\
A  &  \rightarrow SZ^{(\ast)}. \label{Ad}%
\end{align}
One can thus imagine various decay chains, with final states containing
several leptons, jets and missing transverse energy.

For the purposes of detection, the events with charged leptons in the final
state seem most promising. In the region preferred by DM, the decay (\ref{Ad})
gives events having the lepton pair invariant mass sharply peaked at low
values, with a cutoff determined by $\Delta m\lesssim10$ GeV. An extra charged
lepton coming from $H$ via (\ref{Hd}) is likely needed to help discriminate
against the SM background. We have estimated the number of the inert particle
pair production events at the LHC with at least 3 charged leptons in the final
states using \textsc{PYTHIA }\cite{PYTHIA}. In the region preferred by DM, the
process (\ref{HS}) has cross section $\sim0.25$ pb, and a branching ratio (BR)
into at least 3 electrons or muons of $\sim1.5\%$. The effective cross section
of signal events with 3 charged leptons is thus estimated as
\begin{equation}
\sigma_{\text{signal}}\approx3.5\text{ fb}%
\end{equation}
The $H^{+}H^{-}$ pair production has cross section about an order of magnitude
smaller because of the higher mass. The dominant irreducible background is
likely to be the $WZ$ pair production with the $W$ decaying into electrons or
muons and the $Z$ into $\tau$-pairs, with the $\tau$'s also decaying into
electrons or muons. We assume that the background from direct decays of the
$Z$ into electrons or muons can be easily eliminated. In this case we estimate
the effective cross section of background events as
\begin{equation}
\sigma_{\text{bg}}\approx20\text{ fb}%
\end{equation}
An integrated luminosity $\mathcal{L}\sim30$ fb$^{-1}$ might therefore allow a
detection of the signal. It would be very interesting to perform a complete
study going beyond these rough estimates. We are aware of the problems that
might arise from other sources of backgrounds, like the production of
$t\bar{t}$ pairs, which has been studied in an analogous supersymmetric
context \cite{mangano}, or the $W \gamma^{\ast}$ production.

\subsubsection{The Higgs width}

The existence of the new states may be inferred indirectly from the increase
of the width of the usual Higgs. The new decay channels are%
\begin{equation}
h\rightarrow SS,AA,H^{+}H^{-}%
\end{equation}
and the resulting increase in the width of $h$ is%
\begin{equation}
\Delta\Gamma=\frac{v^{2}}{16\pi m_{h}}\left[  \lambda_{S}^{2}\left(
1-\frac{4m_{S}^{2}}{m_{h}^{2}}\right)  ^{1/2}+\lambda_{A}^{2}\left(
1-\frac{4m_{A}^{2}}{m_{h}^{2}}\right)  ^{1/2}+2\lambda_{3}^{2}\left(
1-\frac{4m_{H}^{2}}{m_{h}^{2}}\right)  ^{1/2}\right]  \label{incr}%
\end{equation}
where $\lambda_{S,A}$ are given in (\ref{masses}).

\begin{figure}[ptb]
\begin{center}
\includegraphics[width=8cm]{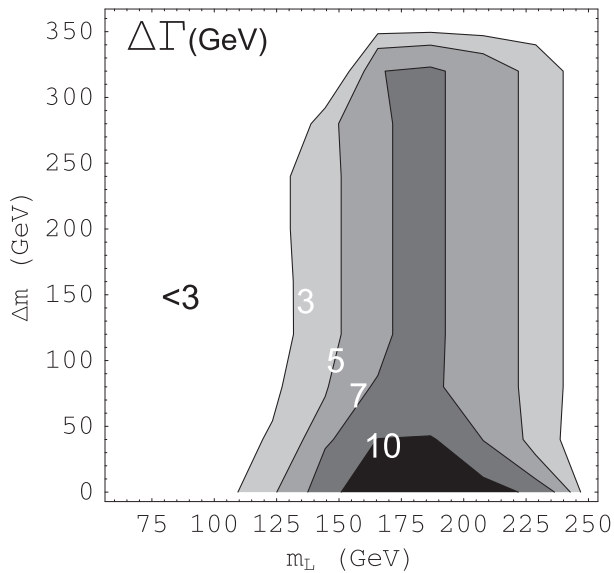}
\end{center}
\caption{{} Contours of the increase in the width of a 500 GeV Higgs,
$\Delta\Gamma$, computed with the maximal couplings allowed by naturalness and
perturbativity.}%
\label{width}%
\end{figure}

The width of a 500 GeV Higgs in the SM is $\Gamma_{SM}\approx68$ GeV
\cite{tome} (mostly due to decays into $WW$,$ZZ$ and $t$\thinspace$\bar{t}$).
If $\Delta\Gamma$ reaches $0.1\,\Gamma_{SM}$, it can be seen with high
luminosity at the LHC. The size of $\Delta\Gamma$ is uncertain, with strong
dependence on $\lambda_{3}$ and on how many channels are open. The maximal
$\Delta\Gamma$ attainable for a given $m_{\text{L }}$and $\Delta m$ is
possible to estimate by letting $\lambda_{\text{L}}$ vary in the range
(\ref{range1}) determined by the naturalness and perturbativity. The resulting
$\Delta\Gamma_{\max}$ is plotted in Fig. \ref{width}. We see that there is a
region where $\Delta\Gamma_{\max}\gtrsim7$ GeV with prospects for the LHC observation.

\section{Alternatives}

\label{alt}Given the indirect nature of the information contained in the EWPT,
there may be and there are in fact other ways to make a heavy Higgs compatible
with them. A discussion of some of the explored possibilities has been given
in Ref. \cite{Peskin}. With reference to this discussion, we point out two
related facts:

1. If the new physics responsible for allowing a heavy Higgs brings in new
4-fermion interactions, it is crucial to check that such new interactions pass
the constraints set by LEP2. A relevant example in this sense is provided by
the way the Kaluza Klein excitations of the SM gauge bosons affect the EWPT.
While their exchange gives effects that indeed allow a good fit of the EWPT
with a heavy Higgs (up to 500 GeV) \cite{StrumiaKK}, this fit becomes
disfavoured by the newer LEP2 data \cite{LEP2}.

2. If one tries a fit of the EWPT with a heavy Higgs, including also the LEP2
data, by adding, one at a time, the four dimension-6 operators involving the
SM Higgs and gauge bosons, but not the fermions, a successful fit is obtained
only from the 4-Higgs operator that corrects $T$ \cite{LEP2}
\begin{equation}
\mathcal{O}_{H}=|H^{\dagger}D_{\mu}H|^{2}.
\end{equation}
This may be of some significance, since this operator is the only one that
breaks custodial symmetry. So it is relatively easier to correct $T$ only, by
adjusting the symmetry breaking parameter(s) that control custodial symmetry,
as we do in the Inert Doublet Model.

As already said, there are other ways of correcting only, or predominantly,
$T$ by some perturbative new physics. As examples, we mention here two possibilities.

i) A scalar triplet of zero hypercharge, $\tau$, \cite{Forshaw:2001xq} coupled
to the SM Higgs via the potential
\begin{equation}
\Delta V=M^{2}\tau^{2}-m\tau_{a}H^{\dagger}\sigma_{a}H.
\end{equation}

ii) A vector-like fermion doublet $F,F^{c}$ of hypercharge 1/2 and a singlet
fermion $S$ with mass and interaction Lagrangian%
\begin{equation}
\Delta\mathcal{L}=\lambda FHS+\lambda^{c}F^{c}H^{\dagger}S+MFF^{c}+\mu S^{2}.
\end{equation}

Both these cases allow to correct predominantly $T$, so that a heavy Higgs
becomes consistent with all current information. In our view, the drawback of
the triplet model is that it corrects $T$ at tree level, so that, depending on
the ratio $m/M$ all the extra particles can be hidden at inaccessible
energies, up to 4-5 TeV. The fermionic example is definitely more constrained
since $T$ receives one loop corrections. It also contains a DM candidate. We
nevertheless find it less compelling than the IDM.

It is in fact natural at this point to ask how the IDM compares with the case
where the second Higgs doublet acquires a non-zero vev. This can also be a way
to improve naturalness \cite{BH}. It should be noted, however, that
distributing the vevs among the 2 doublets, each smaller than $v$, strengthens
the bounds on the cutoff of the loops induced by the quartic self-couplings.
Furthermore, insisting on natural flavour conservation leads to unobserved
massless Goldstone bosons in the limit of exact custodial symmetry. In
particular, to make the charged scalar heavy enough not to conflict with
direct bounds may lead to large contributions to the T parameter. A study of
the consistency of the 2HDM with the EWPT in the space of its parameters has
been discussed in Ref. \cite{Chankowski:2000an}.

\section{Conclusions}

The Large Hadron Collider will explore for the first time an energy domain
well above the Fermi scale. Having in mind that $\Lambda_{QCD}$ is the only
other fundamental scale known in particle physics, the importance of this fact
cannot be overestimated. At the same time we are faced with the success of the
SM, minimally extended to account for neutrino masses, in describing all known
data in particle physics. Which physics will be revealed by the LHC?

Among the many lines of thought that have been followed to try to answer this
question, on one view there is a quite general consensus: the SM is likely to
be the low energy approximation of a more complete theory characterized by one
or more higher physical scales. Is this telling us something about the
complete theory itself? Not the least property of the SM is that its
Lagrangian is the most general renormalizable one for the given gauge symmetry
and particle content. Indeed this apparently allows the SM to be viewed as the
infrared physics of a broad class of theories. All that one needs is to
maintain gauge invariance and to produce a low energy spectrum that matches
the degrees of freedom of the SM. Yet, this last property appears to be
non-trivial due to the presence of the Higgs field. On one side the Higgs is
crucial to the success of the SM in its perturbative description of the data.
On the other side, well identified quantum corrections act to push away the
Higgs from the low energy spectrum of the more complete theory. This is the
naturalness problem of the SM. The problem is particularly compelling in view
of the relatively large numerical size of the relevant quantum corrections,
even cut off at an energy scale well inside the putative range of energies
directly explorable at the LHC. Hence the effort to search for compensating
effects that could be present in the complete theory, of which supersymmetry
is the neatest example. In this view, the LHC will discover the new physics
that cancels the leading quadratic divergences of the SM.

In this paper we have pursued a different line to attack the naturalness
problem of the SM, more modest in scope but physically well motivated, we
believe. The sensitivity of the Higgs mass to the cutoff is after all a
quantitative issue, both for the impact on the physics expected at the LHC and
in connection with the ``LEP paradox'' or the ``little hierarchy
problem''\cite{paradox}. What then if a new physics effect exists which does
not counteract the quadratic divergence of the Higgs mass but nevertheless
relaxes the constraint on the cutoff that is inferred from it? We propose that
such an effect may be due to the presence of a second Higgs doublet which,
however, does not acquire a vev. We find this the simplest way to allow a
heavier mass for the SM Higgs, between 400 and 600 GeV, while keeping full
consistency with the EWPT. In turn a heavier Higgs makes the size of its
quantum corrections less significant: the most important effect is no longer
due to the top loop, as in the unmodified SM, but rather to the loop due to
the Higgs self-coupling. As a consequence, strictly without any cancellation,
the cutoff is pushed to about $1.5$ TeV, against a value of 400 GeV in the SM
with a Higgs mass of 115 GeV. All this happens in a perfectly controllable
perturbative regime for the entire extended model.

The potential of the extended Higgs sector, with a parity symmetry to keep
natural flavour conservation, has 7 parameters, which can be traded for the Z
mass (the vev of the SM Higgs), 4 masses of the scalar particles, 3 neutral
and one charged, and 2 quartic couplings. This potential can support 2
approximate global symmetries: a custodial symmetry, which controls the
splittings among the 3 inert scalars, and a Peccei-Quinn symmetry, which
governs specifically the splitting among the two neutral inert bosons. While
the SM Higgs mass is between 400 and 600 GeV, the other scalars have a mass
ranging from 60 GeV to about 1 TeV. They are always produced in pairs and do
not couple to fermions. It is an interesting question to see if, in the low
mass range, their signals can be seen above background at the LHC.

The lightest of the inert scalars is necessarily stable and is required by
cosmology to be neutral. If the Dark Matter is fully accounted for by this
scalar, its mass is predicted to be around 70 GeV, with a small splitting of
5--10 GeV, controlled by the Peccei-Quinn symmetry, relative to the other
neutral inert scalar, of opposite parity. The pair production of these neutral
bosons may have barely escaped detection at LEP2, due to the small mass
splitting. The cross section on protons of the DM particle is predicted to be
a few times $10^{-9}$ pb, giving a signal below the present limits on direct
DM searches but within the sensitivity of experiments currently under study.

We have stated in the very first paragraph of the Introduction how we view the
status of the EWSB problem in this last year of the pre-LHC era. The
predominant picture, rooted on supersymmetry and theoretically very appealing,
is not without problems. Even more importantly, we find it difficult to say
anything new on it without further experimental inputs. On the other hand we
wonder if alternative roads to LHC physics cannot still be explored. We have
proposed one based on a fully explicit model.

\section*{Acknowledgments}

We would like to thank Alessandro Strumia for many useful conversations, and
Rikard Enberg, Patrick Fox, Gerardo Ganis, Fabiola Gianotti, Jean-Francois
Grivaz, Patric Janot, Tommaso Lari, Michelangelo Mangano, Michele Papucci and
Roberto Tenchini for very useful exchanges concerning Section \ref{coll}. This
work is supported by the EU under RTN contract MRTN-CT-2004-503369. R.B. is
supported in part by MIUR, and L.H. in part by the US Department of Energy
under Contracts DE-AC03-76SF00098, DE-FG03-91ER-40676 and by the National
Science Foundation under grant PHY-00-98840. \appendix

\section{Heavy Higgs RG flow}

\label{SM.pert}

Detailed treatments of the Landau pole constraint in the SM exist
\cite{Hambye}. We will find it instructive to rederive some of the known
results from first principles, focussing on the heavy Higgs case. The one-loop
RG equation for the SM Higgs self-coupling is
\begin{equation}
\frac{d\lambda}{d\ln\Lambda}=\frac{3\lambda^{2}}{2\pi^{2}}+\ldots\text{,}
\label{RG}%
\end{equation}
where \ldots\ stands for the gauge boson and top quark contributions, which
are sub-dominant for heavy Higgs. As discussed below, the appropriate initial
condition for the RG evolution is%
\begin{equation}
\lambda(1.36m_{h})=\frac{m_{h}^{2}}{4v^{2}}\equiv\lambda_{\text{phys}},
\label{init}%
\end{equation}
where the physical Higgs mass $m_{h}$ and its vev $v$ are observable
quantities. The coupling thus evolves as%
\begin{equation}
\lambda(\Lambda)=\frac{\lambda_{\text{phys}}}{1-\frac{3\lambda_{\text{phys}}%
}{2\pi^{2}}\ln\frac{\Lambda}{1.36m_{h}}}\,
\end{equation}
and blows up at the Landau pole%
\begin{equation}
\Lambda_{L}=1.36m_{h}\exp\left(  \frac{2\pi^{2}}{3\lambda_{\text{phys}}%
}\right)  \,.
\end{equation}
In practice, perturbation theory will break down before $\Lambda_{L}$ is
reached. Let us therefore loosely define the perturbativity scale $\Lambda
_{P}$ at which the one-loop correction to $\lambda$ reaches 30\% of the
tree-level value:%
\begin{equation}
\Lambda_{P}=1.36m_{h}\exp\left(  0.3\frac{2\pi^{2}}{3\lambda_{\text{phys}}%
}\right)  \,.
\end{equation}
The values of $\Lambda_{L,P}$ for the Higgs masses in the 400$-$600 GeV range
are given in Table 1 and discussed in Section \ref{SM.p}.

Let us now derive the initial condition (\ref{init}) for the RG evolution
(\ref{RG}). These initial conditions can be read off from the leading
logarithmic dependence of the physical coupling $\lambda_{\text{phys }}$ on
the bare parameters of the Lagrangian, provided that we take care to compute
the precise denominator in the logarithm. We start from the bare Higgs
Lagrangian%
\begin{equation}
L=|\partial H|^{2}-(-\mu_{0}^{2}|H|^{2}+\lambda_{0}|H|^{4}),\quad\lambda
_{0}=\lambda(\Lambda),
\end{equation}
defined with a cutoff $\Lambda$. At the tree level we have%
\begin{equation}
v^{2}=\mu_{0}^{2}/2\lambda_{0},\quad m_{h}^{2}=2\mu_{0}^{2}.
\end{equation}
At the one-loop level the vev should be determined by imposing the vanishing
tadpole condition $\langle h\rangle=0$. The Higgs self-energy gets non-trivial
contributions only from the virtual Higgs pair and Goldstone pair diagrams. We
find the following relation between the (one-loop corrected) vev and the
physical Higgs mass:%
\begin{equation}
\frac{m_{h}^{2}}{4v^{2}}=\lambda_{0}-\frac{3\lambda_{0}^{2}}{2\pi^{2}}\ln
\frac{\Lambda}{C\,m_{h}}\text{,\qquad}C\approx1.36\,. \label{rel}%
\end{equation}
Notice that the coefficient of the logarithm agrees with (\ref{RG}), as it
should. Since the self-energy correction is evaluated at the external momentum
$p^{2}=m_{h}^{2}$, it come as no suprise that $m_{h}$ appears in the
denominator; the exact coefficient 1.36 is found by keeping track of finite
terms. The initial condition (\ref{init}) follows immediately, since the
correction vanishes precisely at $\Lambda=1.36m_{h}$.

\section{2HDM renormalization group equations}

\label{2HDM.rg}The one-loop renormalization group equations of the two-Higgs
doublet model, referred to in Section \ref{our.pert}, are:%
\begin{align}
&  16\pi^{2}\frac{d\lambda_{i}}{d\log\Lambda}=\beta_{i}(\lambda)\nonumber\\
\beta_{1}  &  =24\lambda_{1}^{2}+2\lambda_{3}^{2}+2\lambda_{3}\lambda
_{4}+\lambda_{4}^{2}+\lambda_{5}^{2}\nonumber\\
\beta_{2}  &  =24\lambda_{2}^{2}+2\lambda_{3}^{2}+2\lambda_{3}\lambda
_{4}+\lambda_{4}^{2}+\lambda_{5}^{2}\nonumber\\
\beta_{3}  &  =(12\lambda_{3}+4\lambda_{4})(\lambda_{1}+\lambda_{2}%
)+4\lambda_{3}^{2}+2\lambda_{4}^{2}+2\lambda_{5}^{2}\\
\beta_{4}  &  =4\lambda_{4}(\lambda_{1}+\lambda_{2})+4\lambda_{4}^{2}%
+8\lambda_{3}\lambda_{4}+8\lambda_{5}^{2}\nonumber\\
\beta_{5}  &  =4\lambda_{5}(\lambda_{1}+\lambda_{2})+8\lambda_{3}\lambda
_{5}+12\lambda_{4}\lambda_{5}.\nonumber
\end{align}

\section{Inert doublet contributions to $S,T$}

\label{App.EWPT}We will derive one-loop EWPT corrections induced by the inert
doublet. The $\Delta\rho$ is easiest to compute by relating it to the
wave-function renormalization of the Goldstones $\phi^{+}$ and $\chi$ induced
by the presence of new particles\cite{Barbieri:1992dq}:%
\begin{equation}
\Delta\rho=\delta Z_{\phi}-\delta Z_{\chi}\text{.} \label{deltarho}%
\end{equation}
The relevant cubic interaction Lagrangian between the Goldstones and the inert
particles is the last line of Eq. (\ref{rel1}) below. Goldstone self-energies
get corrected by the diagrams with virtual inert particle pairs. We find:%
\begin{align}
\Delta\rho &  =(\lambda_{4}+\lambda_{5})^{2}f(m_{_{H}},m_{S})+(\lambda
_{4}-\lambda_{5})^{2}f(m_{_{H}},m_{A})-4\lambda_{5}^{2}f(m_{H},m_{S})\\
f(m_{1},m_{2})  &  =\frac{v^{2}}{32\pi^{2}}\int_{0}^{1}\frac{dx\,x(1-x)}%
{xm_{1}^{2}+(1-x)m_{2}^{2}}.
\end{align}
Using (\ref{masses}), it is not difficult to show that this expression is
equivalent to (\ref{extra}).

To find $\Delta S$, we look at the gauge boson self-energy correction
$\Delta\Pi_{BW^{3}}$ due to the virtual $H^{+}H^{-}$ and $SA$ loops. We find:%
\begin{equation}
\Delta S=\frac{1}{2\pi}\int dx\,x(1-x)\ln\frac{xm_{S}^{2}+(1-x)m_{A}^{2}%
}{m_{H}^{2}}.\label{dS}%
\end{equation}
This $\Delta S$ is typically small: $|\Delta S|\lesssim0.1$ in the region
satisfying the naturalness and perturbativity constraints (the same region as
used for determining the typical range of $\Delta T,$ Fig.~\ref{range}),
$-0.04\lesssim\Delta S\lesssim-0.01$ if the $\Delta T$ constraint (\ref{pred})
is imposed. Thus it has no significant effect on the EWPT fit.

\section{Dark Matter (co)annihilation cross sections}

\subsection{Annihilation into gauge bosons}

\label{DM.1}This process is dominant above the $WW$ threshold. Since the
resulting DM abundance will be very small, we will be content with a rough
estimate of the cross section. In particular, we will compute the annihilation
amplitudes in the massless final state approximation. This will be accurate
for $m\gg m_{W}$, and will provide an order-of-magnitude estimate otherwise.
The threshold behavior can be approximated by multiplying with phase space
suppression factors.

We consider annihilation into transverse and longitudinal states separately.
For transverse final states the amplitude is due to the contact term
interactions:%
\begin{equation}
(\sigma_{\text{LL}\rightarrow\perp\perp})v_{\text{rel}}\approx\frac{g^{4}%
}{64\pi m_{\text{L}}^{2}}\left(  2+1/c^{4}\right)  \approx130\text{ pb
}\left(  100\text{ GeV}/m_{\text{L}}\right)  ^{2}. \label{st}%
\end{equation}
Annihilation into longitudinal states can be approximated by annihilation into
massless Goldstones. The relevant terms in the expansion of (\ref{pot}) are%
\begin{align}
V  &  \supset\frac{1}{4}(A^{2}+S^{2})\bigl[2\lambda_{3}\phi^{+}\phi
^{-}+(\lambda_{3}+\lambda_{4})\chi^{2}\bigr]+\frac{\lambda_{5}}{4}(A^{2}%
-S^{2})\chi^{2}\nonumber\\
&  +\frac{v}{\sqrt{2}}\bigl[2\lambda_{1}\left(  2\phi^{+}\phi^{-}+\chi
^{2}\right)  +\lambda_{A}A^{2}+\lambda_{S}S^{2}\bigr]h\label{rel1}\\
&  +\frac{v}{\sqrt{2}}\Bigl\{\bigl[(\lambda_{4}+\lambda_{5})S+i(\lambda
_{4}-\lambda_{5})A\bigr]H^{-}\phi^{+}+\text{c.c.}\Bigr\}+\sqrt{2}v\lambda
_{5}SA\chi\text{.}\nonumber
\end{align}
We find:%
\begin{align}
\mathcal{M}_{SS,AA\rightarrow\chi\chi}  &  =\lambda_{A,S}+\frac{\lambda
_{S,A}m_{h}^{2}}{s-m_{h}^{2}}+2\lambda_{5}^{2}v^{2}\left(  \frac{1}%
{t-m_{A,S}^{2}}+\frac{1}{u-m_{A,S}^{2}}\right) \\
\mathcal{M}_{SS,AA\rightarrow\phi^{+}\phi^{-}}  &  =\lambda_{3}+\frac
{\lambda_{S,A}m_{h}^{2}}{s-m_{h}^{2}}+\frac{(\lambda_{4}\pm\lambda_{5}%
)^{2}v^{2}}{2}\left(  \frac{1}{t-m_{H}^{2}}+\frac{1}{u-m_{H}^{2}}\right) \\
\mathcal{M}_{SA\rightarrow\phi^{+}\phi^{-}}  &  =i\frac{(\lambda_{4}%
^{2}-\lambda_{5}^{2})v^{2}}{2}\left(  \frac{1}{t-m_{H}^{2}}-\frac{1}%
{u-m_{H}^{2}}\right)  .
\end{align}
At freezeout we can neglect $t,u$ compared to $m_{I}^{2}$; in particular,
coannihilations are suppressed. The LIP annihilation amplitudes can be written
as%
\begin{align}
\mathcal{M}_{\text{LL}\rightarrow\chi\chi}  &  \approx\frac{\lambda_{\text{L}%
}s}{s-m_{h}^{2}}+2|\lambda_{5}|-\frac{4\lambda_{5}^{2}v^{2}}{m_{\text{NL}}%
^{2}}\nonumber\\
\mathcal{M}_{\text{LL}\rightarrow\phi^{+}\phi^{-}}  &  \approx\frac
{\lambda_{\text{L}}s}{s-m_{h}^{2}}+|\lambda_{4}|+|\lambda_{5}|-\frac
{(|\lambda_{4}|+|\lambda_{5}|)^{2}v^{2}}{m_{H}^{2}}. \label{amp1}%
\end{align}
We see that these amplitudes depend on $\lambda_{\text{L}}$, which can vary in
a certain range (see Section \ref{summary}). Because of this it can happen
that one of the two amplitudes (\ref{amp1}) is small, but not both. Indeed,
the total annihilation cross section into longitudinal states can be bounded
from below in a $\lambda_{\text{L}}$-independent way as follows:%
\begin{align}
(\sigma_{\text{LL}\rightarrow\Vert\Vert})v_{\text{rel}}  &  \approx\frac
{1}{64\pi m_{\text{L}}^{2}}(|\mathcal{M}_{\text{LL}\rightarrow\chi\chi}%
|^{2}+2|\mathcal{M}_{\text{LL}\rightarrow\phi^{+}\phi^{-}}|^{2})\\
&  \geq\frac{1}{64\pi m_{\text{L}}^{2}}\frac{2}{3}[\mathcal{M}_{\text{LL}%
\rightarrow\chi\chi}-\mathcal{M}_{\text{LL}\rightarrow\phi^{+}\phi^{-}}]^{2}\\
&  =\frac{1}{96\pi m_{\text{L}}^{2}}\left(  |\lambda_{4}|-|\lambda_{5}%
|-\frac{(|\lambda_{4}|+|\lambda_{5}|)^{2}v^{2}}{m_{H}^{2}}+\frac{4\lambda
_{5}^{2}v^{2}}{m_{\text{NL}}^{2}}\right)  ^{2}. \label{slest}%
\end{align}
We have studied the last expression (which in most cases will be an
underestimate) in the typical range of masses $m_{\text{L}}$, $m_{\text{NL }}$
described in Section \ref{summary} and found that it gives a numerical lower
bound:%
\begin{equation}
(\sigma_{\text{LL}\rightarrow\perp\perp}+\sigma_{\text{LL}\rightarrow
\Vert\Vert})v_{\text{rel}}\gtrsim\min[130\text{ pb }\left(  100\text{
GeV}/m_{\text{L}}\right)  ^{2},10\text{ pb], \quad for}\;\;\;m_{\text{L}%
}=(100\text{--}800)\text{ GeV.} \label{sl}%
\end{equation}
The important point is that the bound (\ref{slest}) is \textit{increasing}
with $m_{\text{L}}$, because the growth of the couplings compensates for the
$m_{\text{L}}^{-2}$ suppression. As a result, the sum of (\ref{st}) and
(\ref{slest}) is above 10 pb in the whole range of $m_{\text{L}}$.

\subsection{Co-annihilation into fermions}

\label{DM.2} Below the $WW$ threshold, the \textit{p}-wave suppressed process
$SA \rightarrow Z^{*} \rightarrow\bar{f}f$ is dominant. The cross section is%
\begin{align}
\sigma v_{\text{rel}}  &  =bv_{\text{rel}}^{2}\\
b  &  =\left(  \frac{g}{2c_{\text{w}}}\right)  ^{4}\frac{\sum(g_{V}^{2}%
+g_{A}^{2})}{96\pi m_{\text{L}}^{2}[1-m_{Z}^{2}/(4m_{\text{L}}^{2})]^{2}%
}\text{,}%
\end{align}
where the sum is over all SM fermions, $f$, except for the top quark, and
$\Delta m \ll m_{L}$. In the range of interest, we have
\begin{equation}
b\approx(250 \text{--}60)\text{ pb},\qquad m_{\text{L}}=(60\text{--}80)\text{
GeV.}%
\end{equation}
For $\Delta m < T$, the thermally averaged cross section which enters the
Boltzmann equation is $\langle\sigma v_{\text{rel}}\rangle=6b/x,$ $x=m/T$.

\end{document}